# Proposal For Neuromorphic Hardware Using Spin Devices

Mrigank` Sharad, Charles Augustine, Georgios Panagopoulos, Kaushik Roy
[1]Department of Electrical and Computer Engineering, Purdue University, West Lafayette, IN, USA
msharad@.purdue.edu

**Abstract:** We present a design-scheme for ultra-low power neuromorphic hardware using emerging spin-devices. We propose device models for 'neuron', based on lateral spin valves and domain wall magnets that can operate at ultra-low terminal voltage of ~20 mV, resulting in small computation energy. Magnetic tunnel junctions are employed for interfacing the spin-neurons with charge-based devices like CMOS, for large-scale networks. Device-circuit co-simulation-framework is used for simulating such hybrid designs, in order to evaluate system-level performance. We present the design of different classes of neuromorphic architectures using the proposed scheme that can be suitable for different applications like, analog-data-sensing, data-conversion, cognitive-computing, associative memory, programmable-logic and analog and digital signal processing. We show that the spin-based neuromorphic designs can achieve 15X-300X lower computation energy for these applications; as compared to state of art CMOS designs.

## I. INTRODUCTION

Neural-networks (NN) constitute a powerful computation paradigm that can algorithmically outperform Von-Neumann schemes in numerous data-processing applications [1]-[8]. However, CMOS based hardware implementations of neuromorphic architectures prove inefficient in terms of power consumption and area-complexity. On one hand, digital designs consume large amount of area, whereas, on the other hand, analog designs, although compact, lead to power hungry solutions. This has limited the scope of neural networks to algorithms and software.

In order to tap the potential of neuromorphic computation at the hardware level, the device-circuit models for the neuron and the synapse, apart from being compact, should also achieve low power consumption. In this work we propose the application of spin-devices in NN hardware design that can help achieve these goals.

Ultra low voltage, current-mode operation of magneto-metallic devices like LSV's and DWM's can be used to realize analog summation/integration and thresholding operations, and, can be used to model energy efficient "neurons" [1]-[8]. Such compact, low-resistance, magneto-metallic devices can perform analog-mode-computation, while operating at ultra-low magnitude, pulsed voltage-supply, thereby simultaneously achieving low power consumption as well as small area. We use magnetic-tunnel-junctions (MTJ) to interface the proposed device models for neuron with CMOS, in order to realize different classes of neuromorphic architectures, dedicated to different applications.

In brief, we propose an entirely novel hardware-design scheme which exploits specific spin-device characteristics to perform ultra low energy neuromorphic computation. The presented work involves innovation in device-modeling as well as in the associated circuit-design. It also addresses the architecture level issues related to such a heterogeneous integration, in order to arrive at a comprehensive design solution.

Rest of the paper is organized as follows. A brief introduction to spin torque devices and their application in logic design, proposed in literature, is provided in section 2. Section 3 describes the spin-based device models for neurons, proposed in this work. Neuromorphic circuit design scheme using the proposed devices is described in section 4. Section 5 presents some examples of neuromorphic architectures based on the proposed scheme. The performance and prospects of the proposed design scheme is discussed in section 6. Finally section 7 concludes the paper.

## II. COMPUTING WITH SPIN DEVICES

Recent experiments on spin torque in device structures like lateral spin valve (LSV) [9], [10], domain wall magnets (DWM) [11], [12], and magnetic tunnel junctions, have opened new avenues for spin based computation. Several logic schemes have been proposed using such devices. Hybrid design schemes using MTJ have been explored that aim to club memory with logic and can possibly benefit from reduced memory-data traffic [18]. Use of spin-torque in LSV's facilitated higher degree of spin current manipulation for logic. All spin logic (ASL) proposed in [13], employs cascaded LSV's interacting through spin torque, to realize logic gates and larger blocks like compact full adders [14], based on spin majority evaluation. A number of logic schemes have been proposed based on current driven domain wall motion in magnetic nano-strips [15], [16]. Recently it has been shown that domain wall motion can be achieved with relatively small current density ($10^7$ A/cm$^2$) in magnetic nano-strips with perpendicular magnetic anisotropy [20]. This phenomenon was exploited in a recent proposal on DWM based logic scheme that employed short magnetic nano-wires to model logic gates [15].

Most of the spin based computation schemes proposed so far have been centered on modeling digital logic gates using these devices. A wider perspective on application of spin torque devices however, would involve, not only exploring possible combination of spin and charge devices but, searching for computation models which can derive maximum benefits from such heterogeneous integration.

We noted that ultra low voltage, current-mode operation of magneto-metallic devices like LSV's and DWM's can be used to realize analog summation/integration and thresholding operations with the help of appropriate circuits, and, can be used to model energy efficient "neurons" [1]-[8]. Such device-circuit co-design can lead to ultra low power neuromorphic computation architectures, suitable for different data processing applications. The proposed hybrid design scheme can open a new frontier for spin torque based analog and digital computing.

## III. SPIN BASED DEVICE MODELS FOR NEURON

In this section we present different models for neurons based on spin-devices. Device-models for 'summing – neurons' are discussed in detail. In such a neuron, all input signals are clock synchronized and concurrent. Hence the 'integration' operation in the 'integrate and fire' functionality of a neuron can be simply replaced by summation. A brief description of DWM based 'integrating-neuron' is presented towards the end.

### A. Neuron Models Based on LSV

#### 1. Bipolar Spin Neuron

Fig. 1 shows the device structure for biopolar spin neuron [2], [4], [5], [7]. It constitutes of an output magnet $m_1$ with MTJ based read-port (using a reference magnet $m_5$), and two anti-parallel input magnets $m_2$ and $m_3$, with their 'easy-axis' parallel to that of $m_1$. A preset-magnet $m_4$, with an orthogonal easy-axis, is used to implement current-mode Bennett-clocking (BC) [13].

A current pulse input through $m_4$, presets the output magnet, $m_1$, along its hard-axis. The preset pulse is overlapped with the synchronous input current pulses received through the magnets $m_2$ and $m_3$. After removal of the preset pulse, $m_1$ switches back to its easy-axis. The final spin-polarity of $m_1$ depends upon the sign of the difference $\Delta I$, between the current inputs through $m_2$ and $m_3$. The lower limit on the magnitude of $\Delta I$ (hence, on current per-input for the neuron), for deterministic switching, is imposed by the thermal-noise in the output magnet, and, imprecision in Bennett-Clocking (BC). The effects of these non-idealities have been included in device simulation (fig. 2).

Transfer-function of an artificial neuron can be expressed as the sign-function of weighted sum of inputs, where the individual weights can be either positive or negative. In the proposed device, the neuron functionality is realized by connecting all the positive-weight inputs (excitatory inputs) to its right-spin input-magnet and vice-versa. The output magnet, in effect, evaluates the sign function with the help of Bennett-clocking, where the right-spin state can be regarded as the 'firing state'

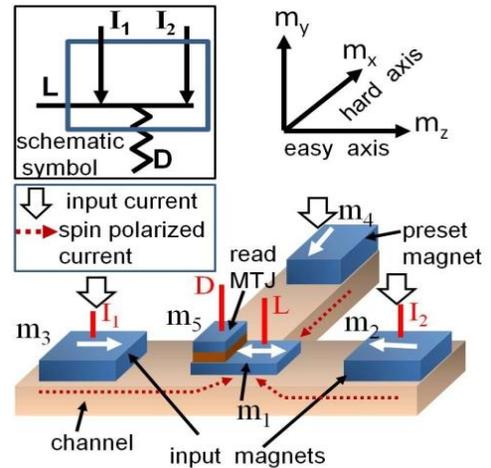

Fig. 1. Bipolar Spin Neuron with local spin injection and decoupled read-write [4].

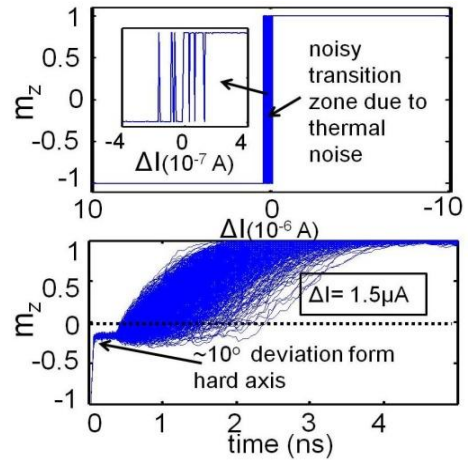

Fig. 2 Due to noise in the neuron-magnet and imprecise BC (leading to $m_z \neq 0$ during preset), larger $\Delta I$ (hence, current for inter-neuron signaling) is required for correct switching, than the ideal case. Minimum inter-neuron signaling current can be determined on the basis of bit-error rate (BER) resulting from these effects.

#### 2. Unipolar Spin Neuron

Fig. 3 shows a slightly different device structure for neuron based on LSV that has a single input magnet. In this case the input magnet receives the difference of current from positive and negative weights magnets, i.e., the subtraction between the two current components is carried out in charge mode, outside the neuron device. As this device receives only the difference $\Delta I$ between the two current components, it can handle larger number of inputs thereby allowing larger scale network. This however comes at the cost of additional circuit design complexity that is discussed later.

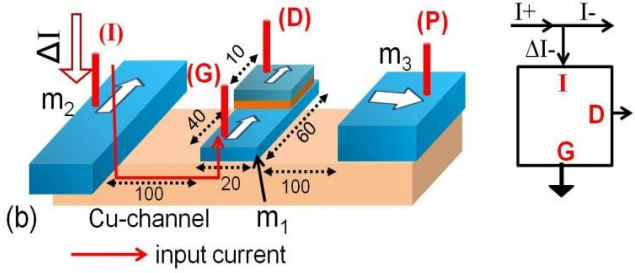

Fig. 3 Unipolar spin neuron.

## 3. *Multi-input spin neuron with DWM synapse*

The device operation explained above can be extended to a multi-input lateral spin valve (LSV) with programmable inputs in the form of DWM (fig. 4a), to realize a compact neuron-synapse unit [1]- [3] (fig, 5).

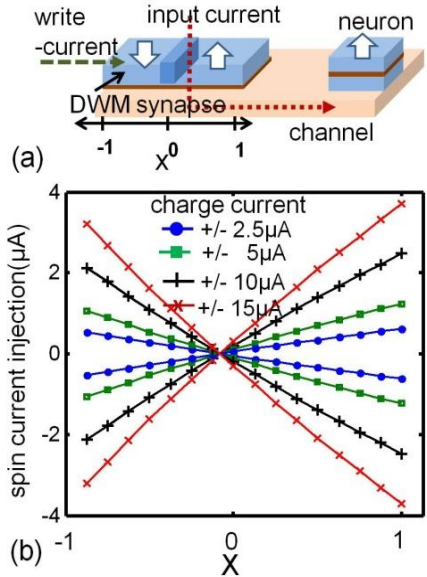

Fig. 4 (a) Domain wall synapse with channel interface (b) Spin polarization strength current injected through DWM as a function of DW location

A DWM constitutes of opposite spin-polarity domains separated by a non-magnetic transition region, termed as the domain wall (DW). The DW can be moved along the nano-magnetic strip by current injection. Hence, a DWM interfaced with the metal channel of an LSV acts as a programmable spin-injector or a spin-mode synapse [1]. The spin-potential in the central region of the channel ( around the ground lead below the output magnet) depends upon the sum of spin currents injected by all the DWM synapses and in turn determines the firing or non-firing state of the neuron, post-Bennett clocking. Fig. 6 depicts the plot for spin-potential in the central region of the channel, surrounding the output magnet of a 16-input neuron, under input conditions corresponding to firing and non-firing conditions. It shows that, in case of a firing event, the entire channel is dominantly at a positive spin potential and vice-versa.

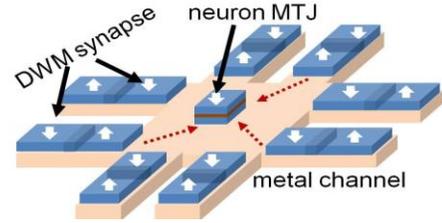

Fig. 5 Spin-based neuron model with three inputs (DWMsynapses). The free layer of the neuron MTJ is in contact with the channel and its polarity, after preset, is determined by spin polarity of combined input current in the channel region (ground terminal) just below it.

In [18] we showed that both local as well as non-local spin torque can be used to realize the neuron models based on LSV described above.

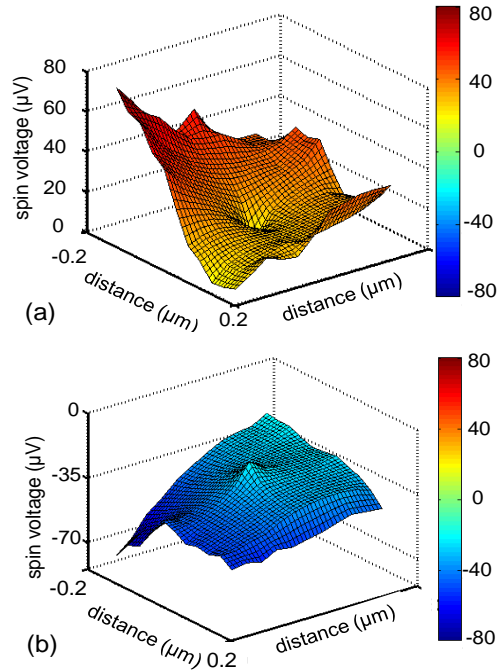

Fig. 6 (a) Channel spin potential of a 16 input neuron under firing condition (b) Channel spin potential under non-firing conidtion.

### B. Neuron Models based on Domain Wall magnet

#### 1. *Unipolar Summing Neuron*

Low current threshold for domain wall motion in Perpendicular Magnetic Anisotropy (PMA) nano-magnet strips [20], can be exploited to model a 'unipolar' neuron shown in fig 7 [6]. It constitutes of a thin and short ($20 \times 60 \times 2$ nm$^3$) DWM nano-strip connecting two antiparallel magnets of fixed polarity, $m_1$ and $m_2$. The magnet $m_1$ forms the input port, whereas, $m_2$ is grounded. Spin-polarity of the DWM layer can be written parallel to $m_1$ or $m_2$ by injecting a small current (~3μA) along it, depending upon the direction current flow [15]. MTJ based detection port is used for reading the spin polarity of the DWM stripe (fig. 7).

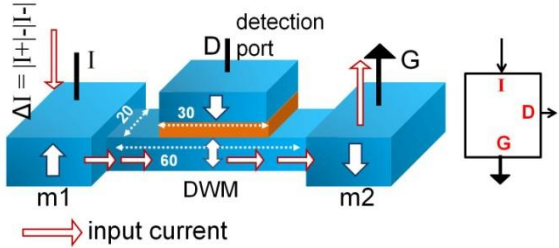

Fig.7 Unipolar spin neuron using domain wall magnet. :

Note that, application of such a structure in memory [20] and digital logic design [15] has been proposed earlier. We exploit this structure to model a neuron using appropriate circuit scheme [6]. The input port of the DWM neuron receives the difference of the positive and the negative synapse currents, $\Delta I$. In addition to this, a bias current can be supplied which effectively shifts the DWM threshold closer to the origin. As a result, a small positive or negative $\Delta I$ (~1µA) can determine evaluation to one of the spin states, thereby realizing the sign function of a neuron. We employ dynamic CMOS latch for reading the MTJ, which results in only a small transient current drawn from the ground terminal (*G*) of the DWM neuron, which can be kept below its switching threshold. Additionally, the time domain threshold for domain wall motion also helps in preventing read disturb from the small transient current [15].

2. *Integrating Neuron Using Domain Wall Magnets*
Spiking neural network is the most recent and evolving topology of neural networks. Among different NN classes, it is regarded as the closest analogue to the biological neural network.

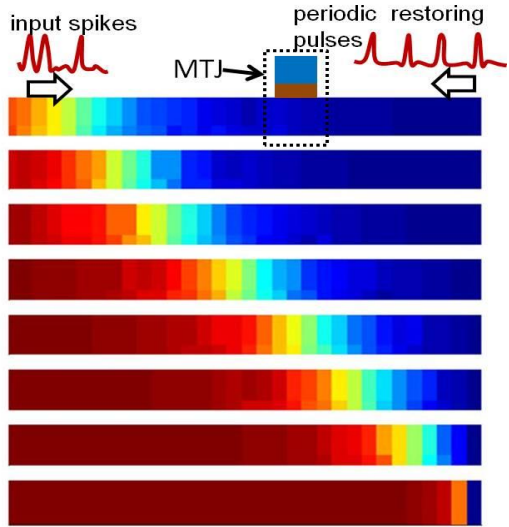

Fig. 8 Integrating neuron using DWM stripe: periodic restoration spikes are used to model 'leaky integration' in the neuron.

It employs asynchronous communication between neurons using spikes. This necessitates time-domain integration of input-signals. Conventionally, dedicated capacitors have been employed for low speed SNN, while analog integrators have been used for getting higher performance. This once again presents the similar bottle neck of area and power consumption as described in the introduction. We propose the use of DWM stripe to realize time domain integration of input spikes. Step-wise motion of domain wall in longer nano-magnet stripes can be used to perform ultra-low voltage current mode integration. Firing state of the neuron can be detected using an MTJ (fig. 8). A DWM based integrating neuron allows spike transmission across ultra low terminal voltage and also mitigates the area overhead of capacitor. Hence it can lead to low power and compact SNN design.

### IV. CIRCUIT INTEGRATION SCHEME

In this section, we describe the circuit integration scheme used in this work that exploits the ultra low voltage operation of the proposed spin neurons for energy efficient, analog-mode neuromorphic computation.

A dynamic CMOS latch senses the state of the neuron MTJ while injecting only a small transient current into the detection terminal [1]. The latch drives transistors operating in deep triode region, which transmit synapse current to all the fan-out neurons (fig. 9). The inter-connection scheme is different for unipolar and bipolar neuron models described in the previous section. For the bipolar neurons, two voltage levels differing by $\Delta V$ are used, i.e., $V$ and $V+\Delta V$ (fig. 9a). Here $V$ is a DC level close to 1V, whereas, $\Delta V$ can be around ~20mV. The source terminal of the output transistors are biased a $V+\Delta V$, where as the ground terminals of the receiving neurons are connected to $V$. Hence, the synapse currents, involved in computation, flow across a small terminal voltage $\Delta V$, thereby, reducing the static power consumption resulting from large number of analog-mode synaptic communications in a neural network

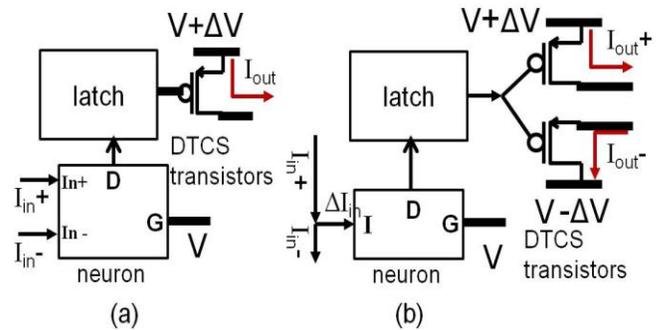

Fig. 9 Circuit integration scheme for (a) bipolar neurons and (b) unipolar neurons. (DTCS: deep triode current source transistos)

For the unipolar neurons, the currents received from negative and positive synapses need to be subtracted in charge mode, outside the device. This necessitates the use of three different voltage levels (fig. 9b). The transistors corresponding to positive weights, effectively source current to the receiving neurons ($I_{out}+$), whereas the transistors corresponding to the negative weights act as drains ($I_{out}-$). In this scheme, most of the current flows between the two extreme levels, $V+\Delta V$ and $V-\Delta V$, whereas, only a small net current flows to and from the mid DC level $V$, through the neuron devices. Hence, routing the additional mid DC level may not be a significant design overhead. However, as the synapse currents in this case flow across $2\Delta V$, for a given strengths of the current source transistors, this scheme leads to 2X higher computation energy as compared to the case of bipolar neuron. Note that, we have chosen two relatively high DC levels differing by $\Delta V$ ($/2\Delta V$), rather than small absolute levels $+/-\Delta V$ ($+/-2\Delta V$), in order to ensure stable supply voltages [1].

## V. DESIGN EXAMPLE

The circuit integration scheme described above can be employed for realizing different classes of neuromorphic architectures. Weights or connection strength between neurons can be realized in different ways. For the multi-input neuron proposed in [1], the DWM inputs act as compact spin-mode synapses (fig. 10).

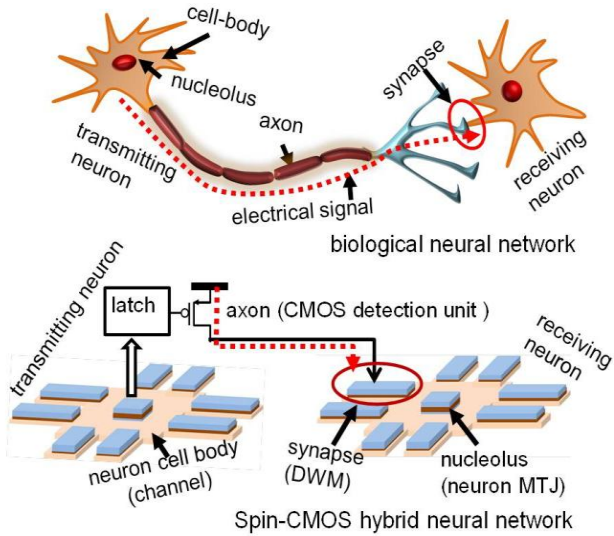

Fig. 10. Correspondence of the spin-CMOS Hybrid ANN to biological neural network: The neuron magnet acts as the firing site, i.e., the nucleolus, the metal channel can be compared to the cell body of the neuron, spin potential in the central region of the channel is analogous to electrochemical potential in the neuron cell body which determines the firing/non-firing state of the neuron, the CMOS detection and transmission unit can be compared to axon of the biological neuron that transmits electrical signal to the receiving neuron, and finally the DWM acts as the synapse.

For other neuron models, weighted source transistors can be used for fixed, non-programmable designs [5]. Fig. 10 depicts a network of DWM neurons, based on this scheme and its analogy to a biological neural network.

Using this technique, we presented the design of an image processing architecture based on cellular neural network (CNN) in [4], [5], [7] (fig. 12). Each neuron in a CNN has two kinds of synaptic connections, type-A and type-B (fig. 12). Through the type-A synapses, a neuron receives the outputs $y_{ij}$, of its eight nearest neighbors and its own state as a feedback. Through type-B synapses, it receives the external signals, $u_{ij}$, (in this work, photo-sensor current from neighboring pixels) from 3x3 surrounding input points.

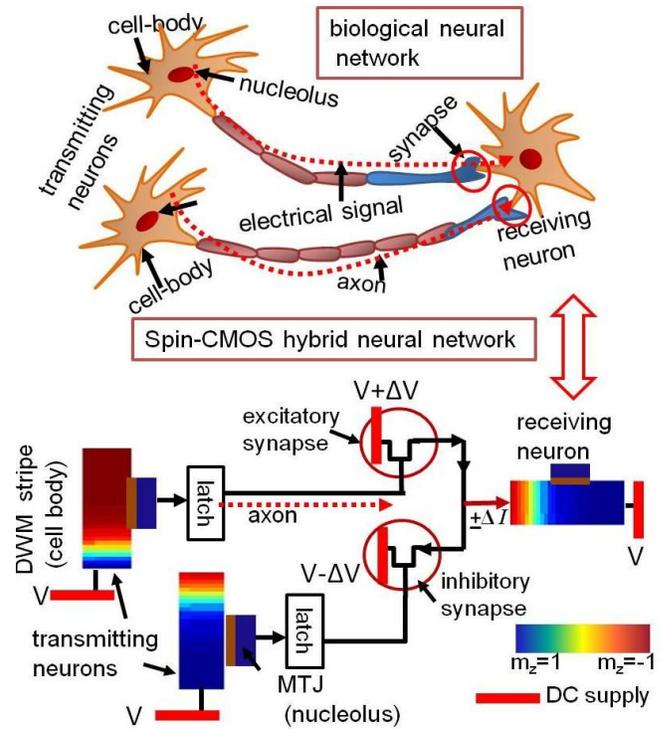

Fig. 11 Emulation of neural network using spin-CMOS hybrid circuit: In each neuron, the MTJ acts as the firing site, i.e., the nucleolus; DWM stripe can be compared to cell body and its spin polarization state is analogous to electrochemical potential in the neuron cell body which affects 'firing', the CMOS detection unit can be compared to axon that transmits electrical signal to the receiving neuron, and finally a weighted transistor acts as synapse as it determines the amount of current injected into a receiving neuron.

The choice of the two sets of weights determine the input-output relation for the whole array and hence the image processing application. The recursive evaluation of neurons in CNN essentially involves weighted sum of these two sets of synaptic inputs, followed by a sign operation

(fig. 12). In the on-sensor image processing architecture presented in [5], *A* and *B*-type synapse weights were realized using weighted triode source transistors, as described above. Note that, in this scheme, the *B*-synapse transistors receive analog-mode photo-sensor voltage at their gate, and, in turn, provide proportional currents to the neurons. On the other hand the *A*-synapse transistors receive binary voltage levels at their gates, corresponding to the source neurons' output state.

Simulation results for some common image processing applications like edge extraction, motion detection, half-toning and digitization (fig. 13), using the spin based CNN, showed ~100x lower computation energy, as compared to state of art mixed-signal CMOS designs. As mentioned earlier, the main advantage comes from ultra low voltage, pulsed operation of spin neurons that are applied to analog computation.

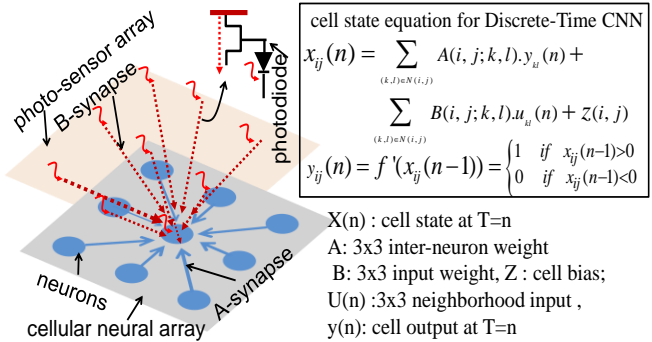

cell state equation for Discrete-Time CNN
$$x_{ij}(n) = \sum_{(k,l) \in N(i,j)} A(i,j;k,l) \cdot y_{kl}(n) + \sum_{(k,l) \in N(i,j)} B(i,j;k,l) \cdot u_{kl}(n) + Z(i,j)$$
$$y_{ij}(n) = f'(x_{ij}(n-1)) = \begin{cases} 1 & \text{if } x_{ij}(n-1) > 0 \\ 0 & \text{if } x_{ij}(n-1) < 0 \end{cases}$$

X(n) : cell state at T=n
A: 3x3 inter-neuron weight
B: 3x3 input weight, Z : cell bias;
U(n) :3x3 neighborhood input ,
y(n): cell output at T=n

Fig. 12. 3x3 neighborhood architecture of CNN and equation for neuron's state: Current from each photosensor $u_{ij}$, is transmitted to 3x3 neighbors through type-*B* synapses implemented using weighted transistors, whereas, inter-neruron connection is determined by type-*A* synapses.

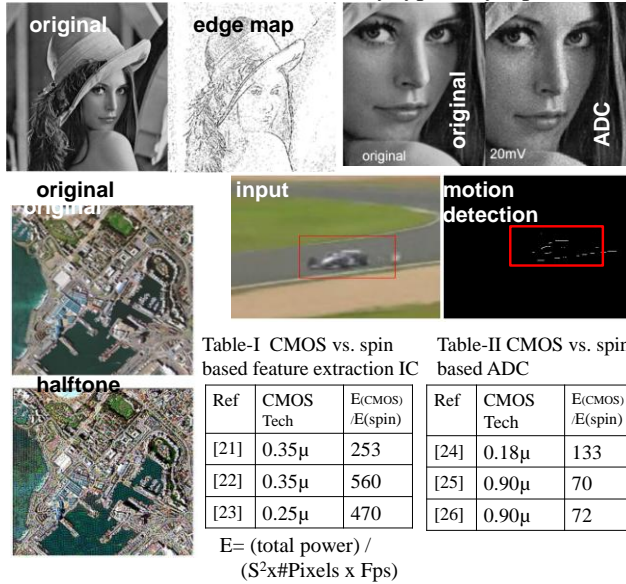

Table-I CMOS vs. spin based feature extraction IC

| Ref | CMOS Tech | $E_{(CMOS)}/E_{(spin)}$ |
|---|---|---|
| [21] | 0.35μ | 253 |
| [22] | 0.35μ | 560 |
| [23] | 0.25μ | 470 |

Table-II CMOS vs. spin based ADC

| Ref | CMOS Tech | $E_{(CMOS)}/E_{(spin)}$ |
|---|---|---|
| [24] | 0.18μ | 133 |
| [25] | 0.90μ | 70 |
| [26] | 0.90μ | 72 |

E= (total power) / ($S^2$x#Pixels x Fps)

Fig. 13 Simulation results for different image processing applications: edge extraction , motion detection , halftoning and digitization; Table 1, 2 compares the energy per computation frame, per pixel, of the propsoed CNN design with some recent CMOS designs for edge extraction and ADC.

Programmable and self-adaptive weights can be realized using programmable conductive elements, like $TiO_2$ memristor or phase change memory (PCM) [2]. Fig. 14 shows a cross-bar neural network architecture using memristor (/PCM) synapses and bipolar spin neurons. Depending upon the polarity of the connectivity between an input line and a neuron, one of the two memristive junctions between them is driven to the off state, while the other is programmed to match the required weight magnitude.

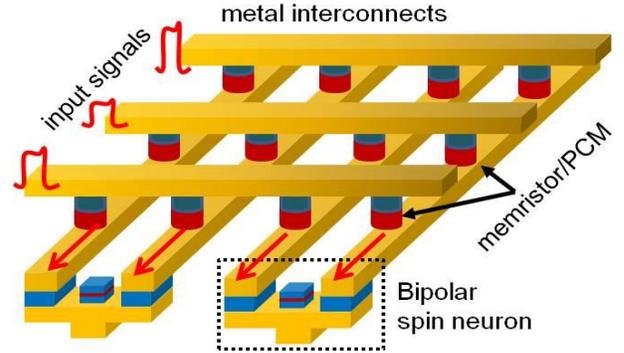

Fig. 14 Cross-bar network design using (a) unipolar spin nueron, (b) using bipolar spin neuron

The spin-neurons facilitate ultra-low voltage, pulsed synaptic communication across the cross-bar metal interconnects, thereby reducing the static-power consumption resulting from large number of inter-neuron signals per-cycle in a large-scale array. Such a design can provide ultra low power solution to several interesting applications, like, logic in memory, associative memory, programmable logic and pattern matching. Spiking neural networks based on memristive cross-bar arrays can realize self-learning networks for cognitive computing. Such a design employs some additional control circuits in each neuron to implement synaptic weight modification according to specific learning rules. But, most of the power consumption in all such networks results from synaptic communication, which can be reduced using DWM based integrating-neurons.

## VI. DESIGN PERFORMANCE

Fig. 15 pictorially depicts the device-circuit co-simulation framework employed in this work to assess the system level performance for different neuromorphic architectures. The device models for neurons have been benchmarked with experimental data on LSVs' and DWM [1]-[7]. The corresponding behavioural models are used for circuit and system level simulation.

Fig. 16 shows the estimated energy benefits of the proposed design scheme over state of art CMOS for different applications. The large benefits for analog applications (2-3 orders of magnitude) comes from the fact that ultra low voltage pulsed operation of spin-based neurons greatly reduce the static power consumption resulting from conventional analog circuits. For applications involving binary signal processing more than 15X-30X lower computation energy has been estimated.

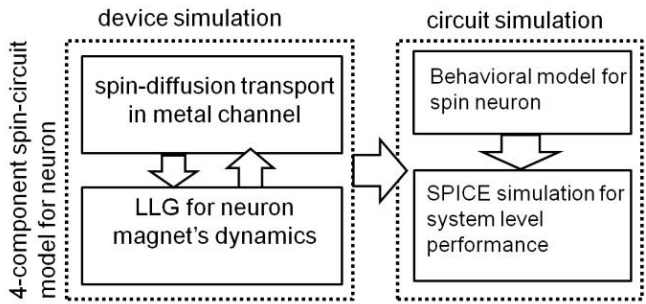

Fig. 15. Device circuit co-simulation framework employed in this work

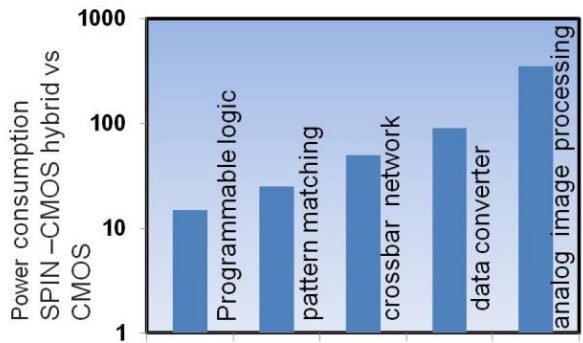

Fig. 16. Energy benefits of the proposed design scheme over CMOS for different applications.

## VII. CONCLUSION

We proposed spin-based device models for neuron that can facilitate the deign of ultra-low power neuromorphic-computation hardware. We developed device-circuit co-simulation framework to assess the performance of heterogeneous neuromorphic designs that employ the proposed neurons. We obtained highly promising estimates for common data processing applications that show 20X-300X improvement in computation energy as compared to state of art CMOS design. The research presented in this work involves device-circuit-architecture co-design and can lead to a comprehensive design solution for neuromorphic hardware.

**ACKNOWLEDGEMENT:** This research was funded in part by Nano Research Initiative and by the INDEX center

**REFERENCE**
[1] M. Sharad et. al, "Spin-based Neuron Model with Domain Wall Magnets as Synapse", IEEE Transaction on Nanotechnology, 2012.
[2] M. Sharad et. al., "Cognitive Computing with Spin Based Neural Networks", DAC 2012
[3] M. Sharad, "Spin-based Neuron-Synapse Units for Ultra Low Power Neural Networks", International Joint Conference on Neural Networks, 2012.
[4] M. Sharad et. al.,"Spin Neurons for Ultra Low Power Computational Hardware", DRC 2012.
[5] M . Sharad et. al., " Ultra Low Energy Analog Image Processing Using Spin Based Neurons", Nanoarch 2012.
[6] M. Sharad et. al., "Boolean and Non-Boolean Computing using Spin Devices", IEDM 2012 (invited).
[7] M. Sharad et. al, arXiv:1205.6022
[8] M . Sharad et. al, arXiv:1206.2466
[9] Kimura et. al., "Switching magnetization of a nanoscale ferromagnetic particle using nonlocal spin injection. Phys. Rev. Lett. 2006
[10]Sun. et. al., "A three-terminal spin-torque-driven magnetic switch", Appl. Phys. Lett. 95, (2009).
[11] M.Yamanouchi et.al., " Velocity of Domain-Wall Motion Induced by Electrical Current in the Ferromagnetic Semiconductor", Physical ReviewLetters, vol.96, pp.096601,2006
[12] D. Chiba, et al., "Control of Multiple Magnetic Domain Walls by Current in a Co/Ni Nano-Wire", Appl. Phys. Express 3, pp. 073004(1-3), 2010.
[13] Behin-Ain et. al., "Proposal for an all-spin logic device with built-in memory", Nature Nanotechnology 2010
[14] C. Augustine et al, "Low-Power Functionality Enhanced Computation Architecture Using Spin-Based Devices", NanoArch, 2011
[15] J. Youngman et al., "Low Energy Magnetic Domain Wall Logic in Short, Narrow, Ferromagnetic Wires", IEEE Mag. Lett., 2012
[16] D. A. Allwood et. al., " Magnetic Domain-Wall Logic ", Science , Vol. 309 no. 5741 pp. 1688-1692 , 2005
[17] S. Matsunaga, et al., "Fabrication of a Nonvolatile Full Adder Based on Logic-in-Memory Architecture Using Magnetic Tunnel Junctions", App. Phy. Exp., 2008.
[18] Fengbo Ren; Markovic, D.; , "True Energy-Performance Analysis of the MTJ-Based Logic-in-Memory Architecture (1-Bit Full Adder),"
[19] D. E. Nikanov et. al., " Uniform Methodology for Benchmarking Beyond CMOS Devices", CRL, Intel cor. , 2012
[20] S. Fukami et al., "Low-current perpendicular domain wall motion cell for scalable high-speed MRAM," VLSI Technology, 2009 Symposium on , vol., no., pp.230-231, 16-18 June 2009
 [21] Jendernalik et al., BPAS, 2011,
[22] Kong et.al, 2007,
[23] Kim et. al., ETRI 2005,
[24] Ozgun et al., ISCAS 2011,
[25] Harpe et. al., ISSCC, 2007,
[26] Craninckx et. al., ISSCC, 2007.